\documentclass[preprintnumbers,superscriptaddress,showkeys,showpacs,byrevtex]{revtex4}
\usepackage{amsmath,amsfonts,amssymb,amscd,amsxtra,amsthm}
\usepackage{graphicx}
\usepackage{epstopdf}
\usepackage{bm}
\begin{document}
\preprint{KIAS-P13016}
\preprint{CYCU-HEP-13-04}
\title{Shear viscosity of quark matter at finite temperature in magnetic fields}
\author{Seung-il Nam}
\email[E-mail: ]{sinam@kias.re.kr}
\affiliation{School of Physics, Korea Institute for Advanced Study (KIAS), Seoul 130-722, Korea}
\author{Chung-Wen Kao}
\email[E-mail: ]{cwkao@cycu.edu.tw}
\affiliation{Department of Physics, Chung-Yuan Christian University (CYCU), Chung-Li 32023, Taiwan}
\date{\today}
\begin{abstract}
We have applied the Green-Kubo formula to investigate the shear viscosity in the SU(2) light-flavor quark matter at finite temperature under the external strong magnetic field $e|\bm{B}|\sim m^2_\pi$. For this purpose, we employ the temperature-modified instanton model and the Schwinger method to induce the magnetic field. The quark spectral function with the finite width  motivated by the instanton model is adopted to compute the shear viscosity. We find that shear viscosity increases as temperature increases even beyond the transition temperature $T_0=170$ MeV if temperature-dependent (TDP) model parameters is used. On the other hand, with temperature-independent ones the shear viscosity starts to drop when temperature goes beyond $T_0$. Although the magnetic field reduces the shear viscosity in terms of the magnetic catalysis, its effects are almost negligible in the chiral-restored phase even for very strong magnetic field, $e|\bm{B}|\approx 10^{20}$ gauss. We also compute the ratio of the shear viscosity and entropy density $\eta/s$. Our numerical results are well compatible with other theoretical results for a wide temperature regions. We obtain the parameterization of the temperature-dependent ratio from our numerical result as ${\eta}/{s}=0.27-{0.87}/{t}+{1.19}/{t^2}-{0.28}/{t^3}$ with $t\equiv T/T_0$ for $T=(100\sim350)$ MeV and $e|\bm{B}|=0$.
\end{abstract}
\maketitle
\section{Introduction}
With the energetic developments of the heavy-ion collision (HIC) experiments by Relativistic Heavy-Ion Collision (RHIC) at BNL and Large Hadron Collider (LHC) at CERN, the properties of quark-gluon plasma (QGP) have been widely investigated. One of the most highlighted observations from the HIC experiments is that QGP behaves as almost perfect fluid, being characterized by the small lower-bound value for its shear viscosity, i.e. the Kovtun-Son-Starinets (KSS) bound~\cite{Kovtun:2004de}: $\frac{\eta}{s}>\frac{1}{4\pi}$. It has also been supported by theories, such as the viscous hydrodynamics~\cite{Song:2010mg} and AdS/QCD~\cite{Policastro:2001yc,Buchel:2003tz}. This observation implies that QGP is a strongly coupled system~\cite{Arsene:2004fa}. The viscous hydrodynamic simulation for the elliptic flow $v_2$ with the Monte-Carlo (MC)-Glauber initial condition reproduced the Au$+$Au collision data with $\frac{\eta}{s}=\frac{1}{4\pi}$, while the value $\frac{\eta}{s}\approx\frac{1}{2\pi}$ did the experimental data with the MC-KLN initial condition~\cite{Abelev:2008ab,Song:2010mg}. These observations indicate that the initial condition for the hydrodynamic simulations is a potential uncertainty to determine the  shear viscosity. It is also worth noting that all the hydrodynamic simulations have employed the temperature-independent shear-viscosity values so far. Hence, the information of the temperature dependency of the shear viscosity is important to estimate the QGP shear viscosity. In addition, the initial quantum fluctuations, such as the color-charge fluctuation, can also cause the uncertainty as well~\cite{Schenke:2012wb}. For a recent status for the  shear viscosity, one may refer to Refs.~\cite{Song:2012ua,Cremonini:2011iq}.

The shear viscosity can be theoretically explored by the Green-Kubo formula in terms of the linear response theory~\cite{Green-Kubo:1957mj,Carrington:1999bw,Fukutome:2007ta,Chen:2006iga,Chen:2009sm,Arnold:2003zc,Arnold:2000dr,Huang:2009ue,Iwasaki:2007iv,Huang:2011ez,Sasaki:2008um}.  Since QGP is a strongly coupled system, it can only be studied via nonperturbative methods in principle, such as low-energy effective QCD-like models or lattice QCD  (LQCD) simulations. From the effective models, such as the Nambu--Jona-Lasinio (NJL) model, the shear viscosity has been scrutinized extensively as a function of temperature ($T$) and/or quark chemical potential ($\mu$)~\cite{Iwasaki:2007iv,Sasaki:2008um,Fukutome:2007ta}. In Refs.~\cite{Fukutome:2007ta,Iwasaki:2007iv}, it was argued that the quark spectral function in a simple local mean-field (MF) only gives trivial results, i.e. $\eta$=0. To overcome this difficulty, one needs to go beyond MF. One of the remedies for this issue is to consider a finite width for the quark spectral function~\cite{Fukutome:2007ta}. The shear viscosity was also investigated by LQCD simulations~\cite{Meyer:2007ic}, dissipative hydrodynamics~\cite{Huang:2009ue,Huang:2011ez}, chiral perturbation theory ($\chi$PT)~\cite{Chen:2006iga},  perturbative QCD  (pQCD)~\cite{Chen:2009sm}, and holographic models~\cite{Policastro:2001yc,Buchel:2003tz,Mamo:2012sy}.

In addition to the shear viscosity of QGP, the effects from the external magnetic field produced in the peripheral HIC experiments have also attracted much attention~\cite{Bzdak:2011yy}. Although the produced magnetic field is reduced by a factor $\sim10^4$ after $\sim3$ fm/c~\cite{Tuchin:2013ie}, its strength is still very strong in the order of pion mass squared: $e|\bm{B}|\propto m^2_\pi\sim10^{18}$ gauss. Such a strength is comparable to the magnetic field of neutron stars. Moreover, recently people have speculated that a strong external magnetic field may generate the chiral magnetic effect and chiral magnetic wave which will induce in the $CP$-odd domains in QGP~\cite{Fukushima:2008xe}. Moreover, the magnetic field to the QCD matter enhances the spontaneous break-down of chiral symmetry (SB$\chi$S), i.e. the magnetic catalysis~\cite{Boomsma:2009yk}. Hence it is interesting to understand how the external magnetic field will change the shear viscosity of QGP.

In the present work, we want to investigate the shear viscosity of the SU(2) light-flavor quark matter at finite temperature under the strong magnetic field. For this purpose, we employ the dilute instanton liquid model (LIM) for the light flavor SU(2) sector~\cite{Schafer:1995pz,Diakonov:2002fq}. This model manifests the nontrivial quark-instanton interactions via the quark zero mode, resulting in the natural UV regulator by construction. Since we are interested in the system at finite temperature, we modified the LIM parameters, such as the average inter-(anti)instanton distance ($\bar{R}$) and  (anti)instanton size ($\bar{\rho}$), using the caloron solution for the Yang-Mills equation with trivial holonomy, i.e. the Harrington-Shepard caloron~\cite{Harrington:1976dj,Diakonov:1988my}. Although we do not have deconfinement order parameters in our present theoretical framework, we consider that the chirally-restored phase can reflect the information of QGP, since the chiral restoration and deconfinement temperatures are almost the same for the system with the finite current-quark mass~\cite{Fukushima:2003fw,Ratti:2005jh,Nam:2009nn}. Since the quark chemical potential is expected to be small inside QGP created in the HIC experiments, we choose $\mu$=0 throughout the present work for brevity. As mentioned above, the Green-Kubo formula is employed to compute the shear viscosity in terms of a quark spectral function~\cite{Fukutome:2007ta}. We construct a quark spectral function with a finite width $\Lambda\sim1/\bar{\rho}$ motivated by the instanton physics. The external magnetic field is induced to the system in hand by the Schwinger method~\cite{Schwinger:1951nm,Nieves:2006xp,Nam:2008ff}. 

Before studying the shear viscosity itself as a function of $T$, we have to investigate the chiral phase transition within the present theoretical framework since the constituent quark mass as a chiral order parameter plays a crucial role in the temperature dependency of the shear viscosity.
By computing $\bar{R}$ and $\bar{\rho}$ as functions of $T$ with the caloron solution, we observe that they are all decrease with respect to temperature, signaling the (partial) restoration of SB$\chi$S~\cite{Diakonov:1988my,Nam:2009nn}. At $T\approx T_0$, where $T_0$ indicate the chiral phase transition temperature, there appear about $10\%$ decreases in $\bar{R}$ and $\bar{\rho}$ in comparison to their values at zero temperature. Using these results and the thermodynamic potential of LIM in the leading $1/N_c$ contributions, we explore the chiral phase transition and shows the second order ($T_0\approx166$ MeV) and crossover ($T_0\approx170$ MeV) transitions for the chiral limit and finite current-quark mass, respectively. It is consistent with the universal restoration pattern. 

The numerical results for the shear viscosity are given as functions of temperature as well as the strength of the external magnetic field with the {\it temperature-dependent parameters}, $\bar{\rho}(T)$ and $\bar{R}(T)$ (TDP), and the {\it temperature-independent parameters}, $\bar{\rho}(0)$ and $\bar{R}(0)$ (TIP). With TIP, the curves for the shear viscosity increase up to $T_0$ then decrease smoothly. On the contrary, the curves keep increasing for TDP beyond $T_0$, signaling the reduction of nonperturbative effects. We also observe a tendency that the external magnetic field lowers the shear viscosity, due to the enhancement of the SB$\chi$S, i.e. the magnetic catalysis via the magnetic field. According to the second-order chiral phase transition in the chiral limit, there is no differences in the shear viscosity due to the magnetic field beyond $T_0$, since the magnetic-field effects are proportional to the constituent-quark mass in the present theoretical framework which is a chiral order parameter. If we go beyond the chiral limit, manifesting the crossover chiral phase transition, we observe finite differences in the shear viscosity curves even for $T>T_0$, but they vanish gradually from $T\approx220$ MeV. In general, the effects from the magnetic field to the shear viscosity are less than $10\%$ for $e|\bm{B}|\lesssim100\times m^2_\pi\approx10^{20}$ gauss.

We also present the ratio of the shear viscosity and the entropy density $\eta/s$ as a function of temperature and the strength of the external magnetic field. We find that $\eta/s$ decreases smoothly and approach the KSS bound for TDP, whereas TIP result undershoots the bound.
Moreover, the effects from the magnetic field becomes almost negligible beyond $T_0$, although the effects are still visible below $T_0$. We also compare the present numerical results for $\eta/s$ with other theoretical estimations from the NJL model, LQCD, and $\chi$PT, resulting in qualitatively good agreement with them. Typical values for the shear viscosity at $T=T_0$ are given as $\eta=0.02\,\mathrm{GeV}^3$ and $\eta/s=0.29$ from the present model for $T=(100\sim350)$ MeV and $n_B=0$.

The present work is organized as follows: In Section II, we provide theoretical framework for computing the shear viscosity of quark matter: the Green-Kubo formula, finite-width quark spectral function, Schwinger method, thermodynamic potential, and so on. The numerical results and relevant discussions are given in Section III and the final Section is devoted for the summary and conclusion.

\section{Theoretical formalism}
In this section we briefly introduce our theoretical framework.
\subsection{Shear viscosity at finite temperature}
First, the static shear viscosity $\eta$ is defined according to the Green-Kubo formula,
~\cite{Fukutome:2007ta}:
\begin{equation}
\label{eq:SV}
\eta=-\frac{\partial}{\partial\omega}\mathrm{Im}[\Pi^\eta_\mathrm{R}(\omega)]|_{\omega=+0},
\end{equation}
where $\Pi^\eta_\mathrm{R}$ stands for the retarded  (R) quark correlation function and $\omega$ for the frequency of the system. The retarded correlation function is related to the correlation function as follows:
\begin{equation}
\label{eq:PI1}
\Pi^{\eta}_\mathrm{R}(\omega)
=\Pi^{\eta}_\mathrm{M}(iw)|_{iw\to \omega+i\epsilon}.
\end{equation}
Here $w$ is the fermionic Matsubara (M) frequency.
$\Pi^{\eta}_\mathrm{M}$ is defined as the time-ordered tensor current correlator,
\begin{equation}
\label{eq:COR3}
\Pi^{\eta}_\mathrm{M}(iw)=-\int^{1/T}_0 d\tau\,e^{-iw\tau}\int\,d\bm{r}
\langle0|{\cal T}\left[J_{xy}(\bm{r},\tau),J_{xy}(0,0) \right]|0\rangle,\,\,\,\,
J_{xy}=\frac{i}{2}\left[\bar{\psi}(\gamma_y\partial_x\psi)-(\partial_x\bar{\psi})\gamma_y\psi \right],
\end{equation}
where $\tau$ and $\psi$ stand for the Euclidean time and quark field. $T$ denotes temperature.
One can evaluate
$\Pi^{\eta}_{\mathrm{M}}$ with the full quark propagator $S$ by using the fermionic Matsubara formula with $w_n=(2n+1)\pi T$:
\begin{eqnarray}
\label{eq:PI2}
\Pi^{\eta}_\mathrm{M}(iw)&=&T\int\frac{d^3\bm{k}}{(2\pi)^3}\sum^\infty_{n=\infty}
\mathrm{Tr}_{c,f,\gamma}\left[k_x\gamma_yS(iw+iw_n,\bm{k})k_x\gamma_yS(iw_n,\bm{k}) \right]
\cr
&=&-\oint\frac{dz}{2\pi i}\int\frac{d^3\bm{k}}{(2\pi)^3}n_F(z)
\mathrm{Tr}_{c,f,\gamma}\left[k_x\gamma_2S(iw+z,\bm{k})k_x\gamma_2S(z,\bm{k}) \right].
\end{eqnarray}
The trace runs over the color ($c$), flavor, ($f$) and Lorentz ($\gamma$) indices. From the first to the second lines in Eq.~(\ref{eq:PI2}), we have employed the fact that the poles of the Fermi-Dirac distribution:
\begin{equation}
\label{eq:FD}
n_F(z)=\frac{1}{1+e^{z/T}},
\end{equation}
are located at $z$=$i(2n+1)\pi T$. The spectral function is related to the quark propagator in the following way,
\begin{equation}
S(k_0,\bm{k})=\int^\infty_{-\infty}
\frac{dw}{2\pi}\frac{\rho(w,\bm{k})}{k_0-w}.
\end{equation}
Hence we obtain the following expression of the shear viscosity in terms of the quark spectral function $\rho(k)$:
\begin{eqnarray}
\label{eq:SHEARV}
\eta&=&-\frac{N_cN_f}{2}\lim_{\omega\to+0}
\int\frac{dk_0}{2\pi}\frac{d^3\bm{k}}{(2\pi)^3}
\frac{[n_F(k_0+\omega)-n_F(k_0)]}
{\omega}k^2_x\mathrm{Tr}_{\gamma}
\left[\rho(k_0+\omega)\gamma_2\rho(k_0)\gamma_2 \right]
\cr
&=&-\frac{N_cN_f}{2}
\int\frac{dk_0}{2\pi}\frac{d^3\bm{k}}{(2\pi)^3}
n'_F(k_0)k^2_x\mathrm{Tr}_{\gamma}
\left[\rho(k_0,\bm{k})\gamma_2\rho(k_0,\bm{k})\gamma_2 \right].
\end{eqnarray}
Here $n'_F$ is defined as,
\begin{equation}
\label{eq:FDD}
n'_F=\frac{\partial n_F(z)}{\partial z}=n_F(z)[1-n_F(z)].
\end{equation}
If we adopt the spectral function associated with the free quark propagator with a constant quark mass $m$ given in Ref.~\cite{Espinosa:2005gq}:
\begin{equation}
\label{eq:PRO}
\rho(w,\bm{k})=2\pi\,\mathrm{sgn}[w]\,(\gamma_0w-\bm{\gamma}\cdot\bm{k}+m)\delta(w^2-\bm{k}^2-m^2).
\end{equation}
(This quark spectral function satisfies the normalization condition $\frac{1}{2\pi}\int\rho(w,\bm{k})dw=\gamma_0$~\cite{Fukutome:2007ta} as shown in Appendix.) Due to the $\delta$-function in the spectral function in Eq.~(\ref{eq:PRO}), the shear viscosity for the free quark at the mean-field level becomes zero, i.e. $\lim_{\epsilon\to0}\int dw f(w)\delta(w+\epsilon)\delta(w)=0$ as long as $f(w)$ is a regular function.
To overcome this problem, we introduce a finite width for the quark spectral function as Refs.~\cite{Fukutome:2007ta,Iwasaki:2007iv}. Thus, we replace the delta function in Eq.~(\ref{eq:PRO}) with a Gaussian functions with a finite width:
\begin{eqnarray}
\label{eq:SPEC}
\delta(w^2-\bm{k}^2-m^2)=\delta(w^2-E^2)&\to&\frac{1}{2\sqrt{2\pi}E_{\bm{k}}\Lambda}\left[\exp\left[-\frac{(w-E_{\bm{k}})^2}{2\Lambda^2} \right]+\exp\left[-\frac{(w+E_{\bm{k}})^2}{2\Lambda^2} \right] \right]\equiv\mathcal{F}(w,\bm{k}),
\end{eqnarray}
where the energy for a quark and the momentum-dependent effective quark mass are defined as in the previous works~\cite{Nam:2012sg} by
\begin{equation}
\label{eq:EM}
E_{\bm{k}}=\sqrt{\bm{k}^2+M_{\bm{k}}^2},\,\,\,\,M_{\bm{k}}=M_0(T)\left[\frac{2}{2+\bar{\rho}^2(T)\,\bm{k}^2} \right]^{2n}.
\end{equation}
Note that the $\Lambda\sim1/\bar{\rho}$ was taken as the width for the Gaussian function. Since the $\bar{\rho}$ is only an interaction range parameter in this model, it is natural to identify it as the finite width for the quark spectral function.

It is worth mentioning that $M_{\bm{k}}$ presents the nonlocal (momentum-dependent) interaction of the quarks. The parameter $n$ in Eq.~(\ref{eq:EM}) will be determined in such a way that $\rho_\mathrm{FW}$ reproduce a typical physical quantity such as the chiral condensate. Note that the constituent-quark mass at zero virtuality $M_0$ and average (anti)instanton size $\bar{\rho}$ are functions of temperature here and will be discussed in detail  below.  Merging Eqs.~(\ref{eq:SPEC}) and (\ref{eq:PRO}), we arrive at the following expression for the finite-width (FW) quark spectral function:
\begin{equation}
\label{eq:RHOREG}
\rho_\mathrm{FW}(w,\bm{k})=2\pi\,\mathrm{sgn}[w]\,(\gamma_0w-\bm{\gamma}\cdot\bm{k}+\bar{M}_{\bm{k}})\mathcal{F}(w,\bm{k}).
\end{equation}
Note that the current quark mass $m$ has been replaced by the momentum-dependent effective quark mass as $m\to \bar{M}_{\bm{k}}$ to regulate the quark-loop integral, in which $\bar{M}_{\bm{k}}$ denotes $M_{\bm{k}}+m$. $\rho_\mathrm{FW}$ also satisfies the normalization condition for the quark spectral function as shown in Appendix.
The chiral condensate can be related to the spectral function in Minkowski space:
\begin{equation}
\label{eq:CC1}
\langle\bar{q}q\rangle=-iN_c\int\frac{d^4p}{(2\pi)^4}\mathrm{Tr}_\gamma[S(p_0,\bm{p})]
=iN_c\int\frac{dw\,d^4p}{(2\pi)^5}\mathrm{Tr}_\gamma\left[\frac{\rho_\mathrm{FW}(w,\bm{p})}{p_0-w} \right].
\end{equation}
Performing the Wick rotation for the temporal direction and integrating over $(w,ik_0, \bm{k})$, one is led to
\begin{equation}
\label{eq:CC2}
\langle\bar{q}q\rangle=-8N_c\int\frac{d^4k}{(2\pi)^4}\int^\infty_0dw
\frac{\bar{M}_{\bm{k}}\mathcal{F}(w,\bm{k})}{k^2_0+w^2}.
\end{equation}
To reproduce the empirical value of the chiral condensate $\langle\bar{q}q\rangle\approx-(250\,\mathrm{MeV})^3$ in the chiral limit~\cite{Diakonov:2002fq}, we choose $n=2$ for Eq.~(\ref{eq:EM}). This choice gives $\langle\bar{q}q\rangle\approx-(239\,\mathrm{MeV})^3$ from Eq.~(\ref{eq:CC2}) numerically which is comparable to the empirical value. Throughout the present work, we will use $n=2$ for all the numerical calculations.

Taking into account all the ingredients discusses so far, we arrive at the following neat expressions for the shear viscosity:
\begin{eqnarray}
\label{eq:FSV}
\eta&=&\frac{N_cN_f}{2\pi^2 T}
\int dk_0\, d^3\bm{k}\,
n_F(k_0)[n_F(k_0)-1]\mathcal{F}^2(w,\bm{k})\,
k^2_x\left[2k^2_y+k^2-M_{\bm{k}}^2\right].
\end{eqnarray}
\subsection{Temperature-dependent quark mass}
Here we would like to address how to determine the temperature dependence of the effective quark mass $M_0$ in Eq.~(\ref{eq:EM}). In Refs.~\cite{Nam:2009nn}, we derived it by using the caloron distribution with trivial holonomy, i.e. Harrington-Shepard caloron~\cite{Harrington:1976dj,Diakonov:1988my}. Firstly, we want to explain briefly how to modify $\bar{\rho}$ and $\bar{R}$ as functions of $T$, using the caloron solution. Details can be found in Ref.~\cite{Nam:2009nn}.  An instanton distribution function for arbitrary $N_c$ and $N_f$ can be written with a Gaussian suppression factor as a function of $T$ and an arbitrary instanton size $\rho$ for pure-glue QCD~\cite{Diakonov:1988my}:
\begin{equation}
\label{eq:IND}
d(\rho,T)=\underbrace{C_{N_c}\,\Lambda^b_{\mathrm{RS}}\,
\hat{\beta}^{N_c}}_\mathcal{C}\,\rho^{b-5}
\exp\left[-(A_{N_c}T^2
+\bar{\beta}\gamma n\bar{\rho}^2)\rho^2 \right].
\end{equation}
We note that the CP-invariant vacuum was taken into account in Eq.~(\ref{eq:IND}), and we assumed the same analytical form of the distribution function for both the instanton and anti-instanton. Note that the instanton number density (packing fraction) $N/V\equiv n\equiv1/\bar{R}^4$ and $\bar{\rho}$ have been taken into account as functions of $T$ implicitly. For simplicity, we take the numbers of the anti-instanton and instanton are the same, i.e. $N_I=N_{\bar{I}}=N$. We also assigned the constant factor in the right-hand-side of the above equation as $\mathcal{C}$ for simplicity. The abbreviated notations are also given as:
\begin{eqnarray}
\label{eq:PARA}
\hat{\beta}&=&-b\ln[\Lambda_\mathrm{RS}\rho_\mathrm{cut}],\,\,\,\,
\bar{\beta}=-b\ln[\Lambda_\mathrm{RS}\langle R\rangle],\,\,\,
C_{N_c}=\frac{4.60\,e^{-1.68\alpha_{\mathrm{RS}} Nc}}{\pi^2(N_c-2)!(N_c-1)!},
\cr
A_{N_c}&=&\frac{1}{3}\left[\frac{11}{6}N_c-1\right]\pi^2,\,\,\,\,
\gamma=\frac{27}{4}\left[\frac{N_c}{N^2_c-1}\right]\pi^2,\,\,\,\,
b=\frac{11N_c-2N_f}{3}.
\end{eqnarray}
Note that we defined the one-loop inverse charges $\hat{\beta}$ and $\bar{\beta}$ at certain phenomenological cutoffs $\rho_\mathrm{cut}$ and $\langle R\rangle\approx\bar{R}$. $\Lambda_{\mathrm{RS}}$ denotes a scale, depending on a renormalization scheme, whereas $V_3$ for the three-dimensional volume. Using the instanton distribution function in Eq.~(\ref{eq:IND}), we can compute the average value of the instanton size $\bar{\rho}^2$ straightforwardly as follows~\cite{Schafer:1996wv}:
\begin{equation}
\label{eq:rho}
\bar{\rho}^2(T)
=\frac{\int d\rho\,\rho^2 d(\rho,T)}{\int d\rho\,d(\rho,T)}
=\frac{\left[A^2_{N_c}T^4
+4\nu\bar{\beta}\gamma n \right]^{\frac{1}{2}}
-A_{N_c}T^2}{2\bar{\beta}\gamma n},
\end{equation}
where $\nu=(b-4)/2$. It can be easily shown that Eq.~(\ref{eq:rho}) satisfies the  following asymptotic behaviors~\cite{Schafer:1996wv}:
\begin{equation}
\label{eq:asym}
\lim_{T\to0}\bar{\rho}^2(T)=\sqrt{\frac{\nu}{\bar{\beta}\gamma n}},
\,\,\,\,
\lim_{T\to\infty}\bar{\rho}^2(T)=\frac{\nu}{A_{N_c}T^2}.
\end{equation}
Here, the second relation of Eq.~(\ref{eq:asym}) indicates a correct scale-temperature behavior at high $T$, i.e., $1/\bar{\rho}\approx\Lambda\propto T$. Substituting Eq.~(\ref{eq:rho}) into Eq.~(\ref{eq:IND}), the caloron distribution function can be evaluated further:
\begin{equation}
\label{eq:dT}
d(\rho,T)=\mathcal{C}\,\rho^{b-5}
\exp\left[-\mathcal{F}(T)\rho^2 \right],\,\,\,\,
\mathcal{F}(T)=\frac{1}{2}A_{N_c}T^2+\left[\frac{1}{4}A^2_{N_c}T^4
+\nu\bar{\beta}\gamma n \right]^{\frac{1}{2}}.
\end{equation}
The instanton packing fraction $n$ can be computed self-consistently, using the following equation:
\begin{equation}
\label{eq:NOVV}
n^\frac{1}{\nu}\mathcal{F}(T)=\left[\mathcal{C}\,\Gamma(\nu) \right]^\frac{1}{\nu},
\end{equation}
where we replaced $NT/V_3\to n$, and $\Gamma(\nu)$ stands for the $\Gamma$-function with an argument $\nu$. Note that $\mathcal{C}$ and $\bar{\beta}$ can be determined easily using Eqs.~(\ref{eq:rho}) and (\ref{eq:NOVV}), incorporating the vacuum values for $n\approx(200\,\mathrm{MeV})^4$ and $\bar{\rho}\approx(600\,\mathrm{MeV})^{-1}$: $\mathcal{C}\approx9.81\times10^{-4}$ and $\bar{\beta}\approx9.19$. Finally, in order for estimating the $T$-dependence of $M_0$, one needs to consider the normalized distribution function, defined as follows,
\begin{equation}
\label{eq:NID}
d_N(\rho,T)=\frac{d(\rho,T)}{\int d\rho\,d(\rho,T)}
=\frac{\rho^{b-5}\mathcal{F}^\nu(T)
\exp\left[-\mathcal{F}(T)\rho^2 \right]}{\Gamma(\nu)}.
\end{equation}
Here, the subscript $N$ denotes the normalized distribution. For brevity, we want to employ the large-$N_c$ limit to simplify the expression for $d_N(\rho,T)$. In this limit, as understood from Eq.~(\ref{eq:NID}), $d_N(\rho,T)$ can be approximated as a $\delta$-function:
\begin{equation}
\label{eq:NID2}
\lim_{N_c\to\infty}d_N(\rho,T)=\delta[{\rho-\bar{\rho}(T)}].
\end{equation}
The numerical result for the trajectory of $\bar{\rho}(T)$ is given in the panel (a) in Figure~\ref{FIG01}. Here we choose $\bar{\rho}(0)=1/\Lambda\approx1/3$ fm and $\bar{R}\approx1$ fm for all the numerical calculations. These values are phenomenologically preferred in the present model~\cite{Diakonov:2002fq}. The curve for $\bar{\rho}(T)$ shows that the average (anti)instanton size smoothly decreases with respect to temperature. It indicates that the instanton ensemble gets diluted and the nonperturbative effects via the quark-instanton interactions are diminished. At $T=(150\sim200)$ MeV, which is close to the chiral phase transition temperature, the instanton size decreases by about $(10\sim20)\%$ in comparison to its value at $T$=0. Considering that the instanton size corresponds to the scale parameter of the model, i.e. UV cutoff mass, $\bar{\rho}\approx1/\Lambda$, the temperature-dependent cutoff mass is a clearly distinctive feature in comparison to other low-energy effective models, such as the NJL model. In addition, we also show the temperature dependence of the average (anti)instanton number density or (anti)instanton packing fraction, $N/V$, in the panel (a) of Figure~\ref{FIG01}. Again, the instanton number density decreases as temperature increases: The instanton ensemble is diluted. We will use these two temperature-dependent quantities for computing the shear viscosity in Eq.~(\ref{eq:FSV}).

\subsection{Shear viscosity under strong external magnetic field}
Here, we briefly discuss how to induce the external magnetic field to the quark matter. Following the Schwinger method, we apply the minimal gauge substitution to the covariant derivative, $i\partial_\mu\to iD_\mu=i\partial_\mu+ie_qA_\mu$. By doing that, the momentum dependent effective quark mass can be expanded in terms of the electric charge of the quark, then we obtain the following expression for $\mathcal{O}(e_q)$~\cite{Nam:2011vn}:
\begin{equation}
\label{eq:MMEE}
M_{\bm{k}}\to M_{\bm{k}}+\frac{i}{2}(\sigma\cdot F)\tilde{M}_{\bm{k}},
\,\,\,\,
\tilde{M}_{\bm{k}}=-\frac{32M_0\bar{\rho}^2}{(2+\bar{\rho}^2\bm{k}^2)^5}
\,\,\,\,\mathrm{for}\,\,\,\,n=2.
\end{equation}
We choose the specific configuration for the external magnetic field for convenience as
\begin{equation}
\label{eq:BB}
\bm{B}=(B_x,B_y,B_z)=(0,B_0\sin\theta_B,B_0\cos\theta_B),
\end{equation}
where $\theta_B$ is an arbitrary angle.
It has been verified that choosing arbitrary field configuration does not generate any qualitative difference.
Considering that $1\,\mathrm{G}=1.95\times10^{-14}\,\mathrm{MeV}^2$ in the natural unit and $m^2_\pi\approx10^{18}\,\mathrm{G}$ in terms of the pion mass $m_\pi\approx140$ MeV, it is quite convenient to employ the following parameterization for the magnetic field: $eB_0=n_B m^2_\pi$. As for $n=1$, the strength of the magnetic field is comparable to that of the magnetar. If $n_B$ becomes about ($10\sim100$), it can be compared to the strong magnetic field observed at the peripheral heavy-ion collisions at RHIC~\cite{Tuchin:2013ie}.

Combining all ingredient we have the simple expression for the shear viscosity as a function of $(T,B_0)$ up to $\mathcal{O}(e^2_q)$:
\begin{eqnarray}
\label{eq:FSV2}
\eta(T,B_0)&=&\sum_{q=u,d}\frac{N_c}{2\pi^2 T}
\int dk_0\, d^3\bm{k}\,n_F(k_0)[n_F(k_0)-1]\mathcal{F}^2(k)\,k^2_x
\left[2k^2_y+k^2-M^2_{\bm{k}}+3(e_qB_0)^2\tilde{M}^2_{\bm{k}}\right],
\end{eqnarray}
where the summation runs over the light flavors $u$ and $d$. Corresponding electrical quark charges are $(e_u,e_d)=(+2/3,-1/3)e$, in which $e$ denotes the unit electrical charge $e=\sqrt{4\pi\alpha_\mathrm{EM}}$ in the natural unit. Note that the magnetic field effect comes only from $3(e_qB_0)^2\tilde{M}^2_{\bm{k}}$ which is only proportional to $M_0$ as understood as in Eq.~(\ref{eq:MMEE}).

\subsection{Calculating Entropy in our model}
\begin{figure}[t]
\begin{tabular}{cc}
\includegraphics[width=8.5cm]{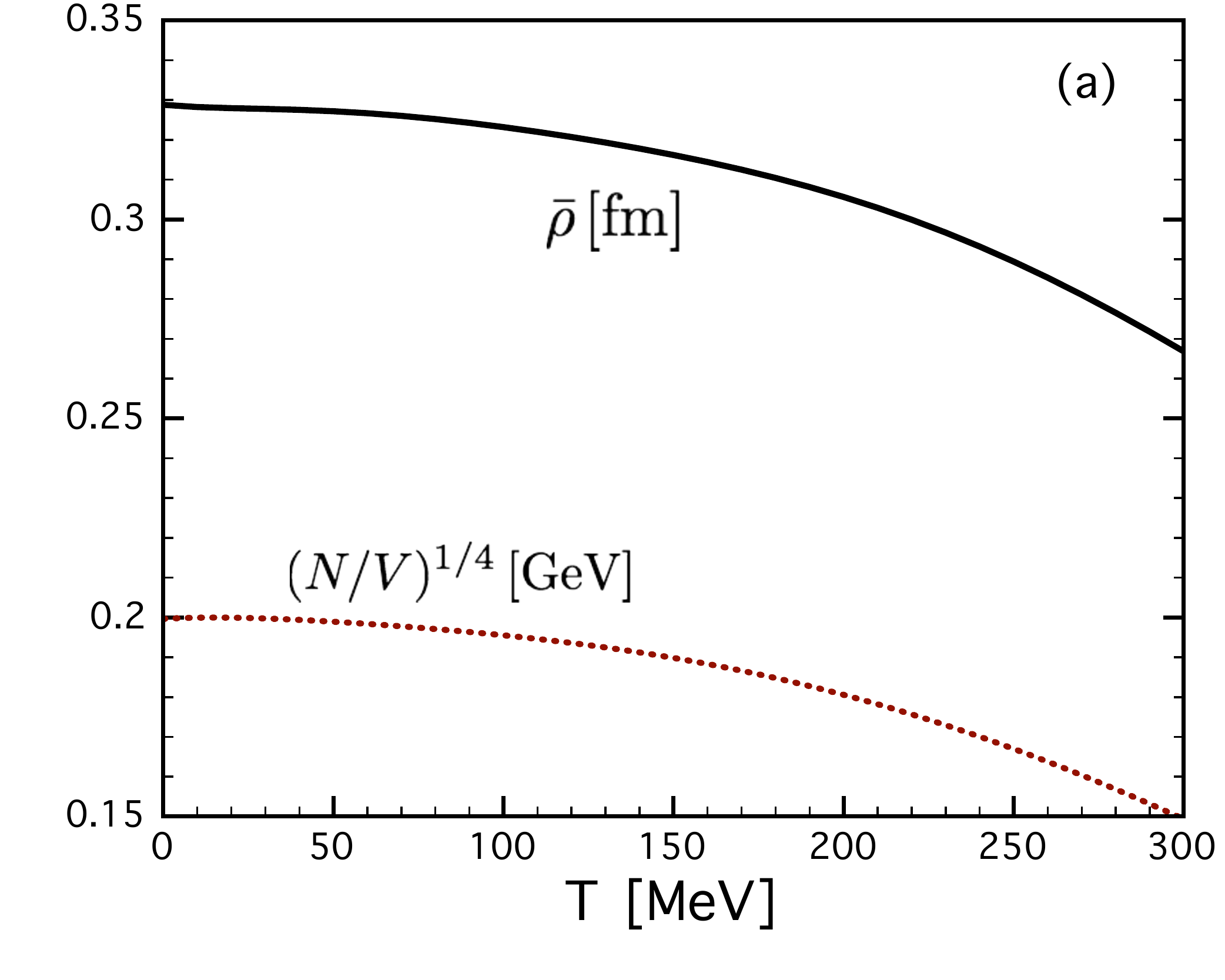}
\includegraphics[width=8.5cm]{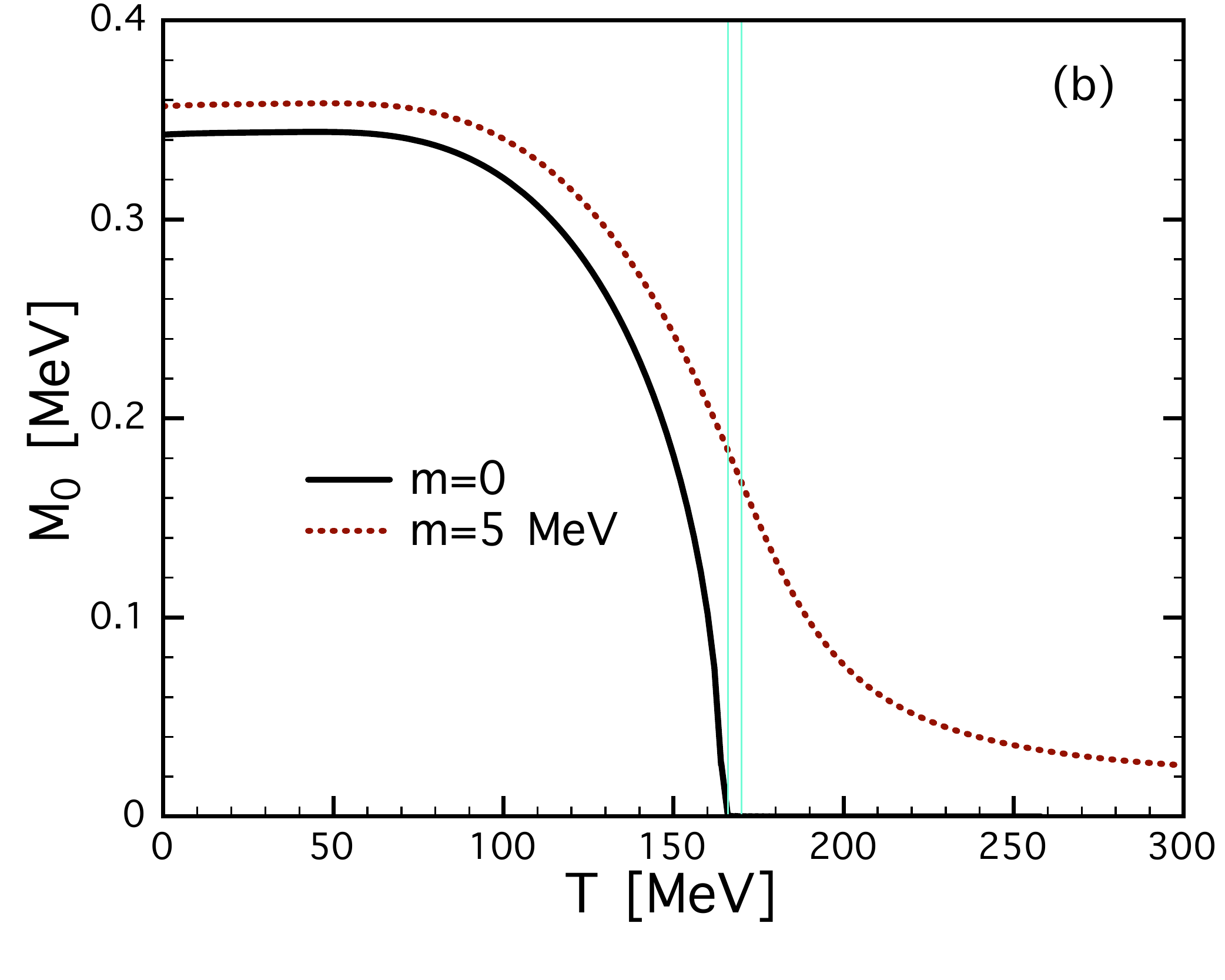}
\end{tabular}
\caption{Average (anti)instanton size $\bar{\rho}\approx1/\Lambda$ [fm] and (anti)instanton packing fraction $(N/V)^{1/4}$ [GeV] as functions of $T$, computed from the Harrington-Shepard caloron distribution~\cite{Harrington:1976dj,Diakonov:1988my} in the panel (a). Effective quark mass at zero virtuality, $M_0$ computed from Eq.~(\ref{eq:LIMGAP}) as functions of $T$ for $m=0$ (solid) and $m=5$ MeV (dot), signaling the second-order and crossover chiral phase transitions, respectively, in the panel (b). The vertical lines indicate the chiral-phase-transition temperatures $T_0=(166,170)$ MeV for $m=(0,5)$ MeV.}
\label{FIG01}
\end{figure}

As in Ref.~\cite{Nam:2009nn}, the LIM thermodynamic potential per volume in the leading $1/N_c$ contributions at zero quark chemical potential can be written as follows:
\begin{eqnarray}
\label{eq:TP}
\Omega_\mathrm{LIM}
&=&\frac{N}{V}\left[1-\ln\frac{N}{\lambda V\mathrm{M}} \right]+2\sigma^2-2N_cN_f\int^\infty_0\frac{d^3\bm{k}}{(2\pi)^3}
\left[E_{\bm{k}}+2T\ln\left[1+e^{-\frac{E_{\bm{k}}}{T}}  \right]
 \right],
\end{eqnarray}
where $\lambda$ and $\mathrm{M}$ represent a Lagrange multiplier to exponentiate the effective quark-instanton action and an arbitrary massive parameter to make the argument for the logarithm dimensionless. $\sigma$ stands for the isosinglet scalar meson field corresponding to the effective quark mass. In the leading $1/N_c$, we have the relation $2\sigma^2=N/V$~\cite{Nam:2009nn}. The gap  equation can be derived from Eq.~(\ref{eq:TP}) by differentiating $\Omega_\mathrm{LIM}$ by the Lagrange multiplier $\lambda$:
\begin{equation}
\label{eq:LIMGAP}
\frac{\partial\Omega_\mathrm{LIM}}{\partial \lambda}=0\to
\frac{N_f}{\bar{M}_0}\frac{N}{V}-2N_cN_f\int^\infty_0\frac{d^3\bm{k}}{(2\pi)^3}
F^4_{\bm{k}}\frac{M_0}{E_{\bm{k}}}\left[1-\frac{2e^{-\frac{E_{\bm{k}}}{T}}}{1+e^{-\frac{E_{\bm{k}}}{T}}}\right]=0.
\end{equation}
Note that one can write the instanton packing fraction in terms of the effective quark mass $M_0$ and $\bar{\rho}$~\cite{Diakonov:2002fq}:
\begin{equation}
\label{eq:NOV}
\frac{N}{V}=\frac{\mathcal{C}_0N_cM^2_0}{\pi^2\bar{\rho}^2}.
\end{equation}
The value of $\mathcal{C}_0$ locates in $(1/3\sim1/4)$ for $1/\bar{\rho}\approx600$ MeV, $M_0\approx(300\sim400)$ MeV and $N/V\approx(200\sim260\,\mathrm{MeV})^4$ for vacuum~\cite{Goeke:2007bj}. We choose $\mathcal{C}_0=0.27$ to reproduce $M_0=(340\sim350)$ MeV at $(T,\mu)=0$ in the chiral limit. After solving Eq.~(\ref{eq:LIMGAP}) with respect to $M_0$ numerically, the numerical results for $M_0$ as a function of $T$ are given in the panel (b) of Figure~\ref{FIG01} for the zero and finite current quark mass: $m=0$ (solid) and $m=5$ MeV (dotted). These results indicate correct universal pattern for the phase transition pattern like the those of the Ising model, i.e. the second-order chiral phase transition for the massless fermion and the crossover for the finite mass. Here we choose the current quark mass to be about $5$ MeV, considering the isospin symmetry for the light SU(2) flavor sector: $m_u\approx m_d\approx m=5$ MeV. From those numerical results, the phase transition temperatures for the two cases are obtained as $T_0\approx(166,170)$ MeV for $m=(0,5)$ MeV. The transition temperatures are indicated by the thin solid vertical lines in the panel (b) of Figure~\ref{FIG01}. Since we are interested in the ratio of the shear viscosity and the entropy density $\eta/s$, we derive the entropy density $s$ as follows:
\begin{equation}
\label{eq:ENT}
s\equiv-\frac{\partial\Omega_\mathrm{LIM}}{\partial T}.
\end{equation}
From the effective thermodynamic potential in  Eq.~(\ref{eq:TP}), we obtain entropy density within the present model:
\begin{equation}
\label{eq:ENT1}
s\approx-\left(\frac{\partial}{\partial T}\frac{N}{V} \right)
\left[1-\ln\left(\frac{N}{V\Lambda^4} \right) \right]+4N_cN_f\int\frac{d^3\bm{k}}{(2\pi)^3}
\left\{\ln\left[1+e^{-E_{\bm{k}}/T} \right]+\frac{E_{\bm{k}}}{T}n_F(E_{\bm{k}}) \right\}.
\end{equation}
In deriving Eq.~(\ref{eq:ENT1}), we assume that $2\sigma^2\approx N/V$ and $\lambda\mathrm{M}\approx\Lambda^4$ as in the leading $1/N_c$, since $\Lambda$ is only the scale parameter of the present model.
The logarithm term $\ln[\cdots]$ in the first square bracket in the right-hand-side of Eq.~(\ref{eq:ENT1}) gives small contribution to the entropy density. As understood in the panel (a) of Figure~\ref{FIG01}, the (anti)instanton number density $N/V$ is a function of temperature, so that its derivative with respect to $T$ in the first them in right-hand-side of Eq.~(\ref{eq:ENT1}) is finite in general within the present model. The detailed calculations for these quantities will be given in a separated work~\cite{sinam}.

\section{Numerical results and Discussions}
\begin{figure}[t]
\begin{tabular}{cc}
\includegraphics[width=8.5cm]{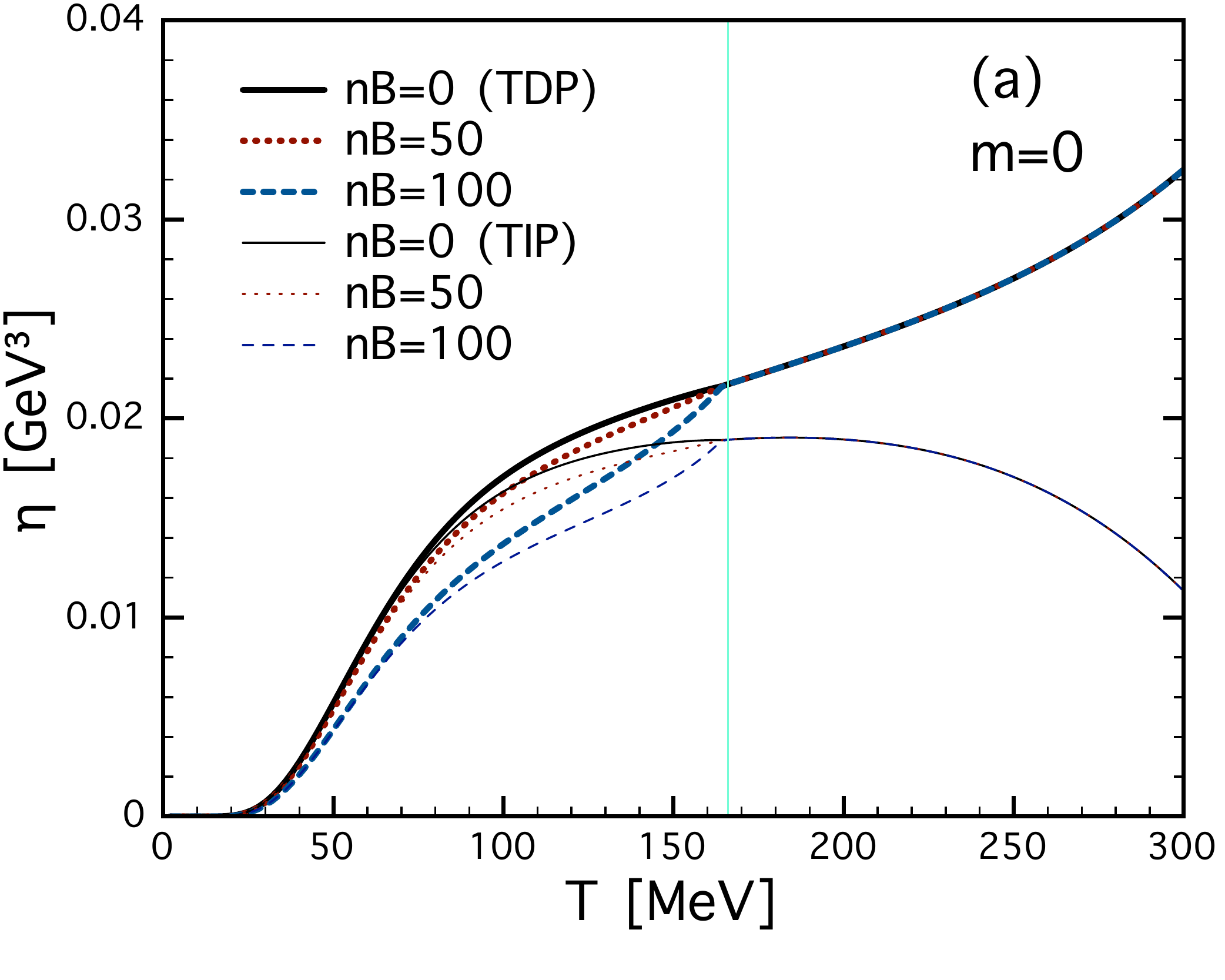}
\includegraphics[width=8.5cm]{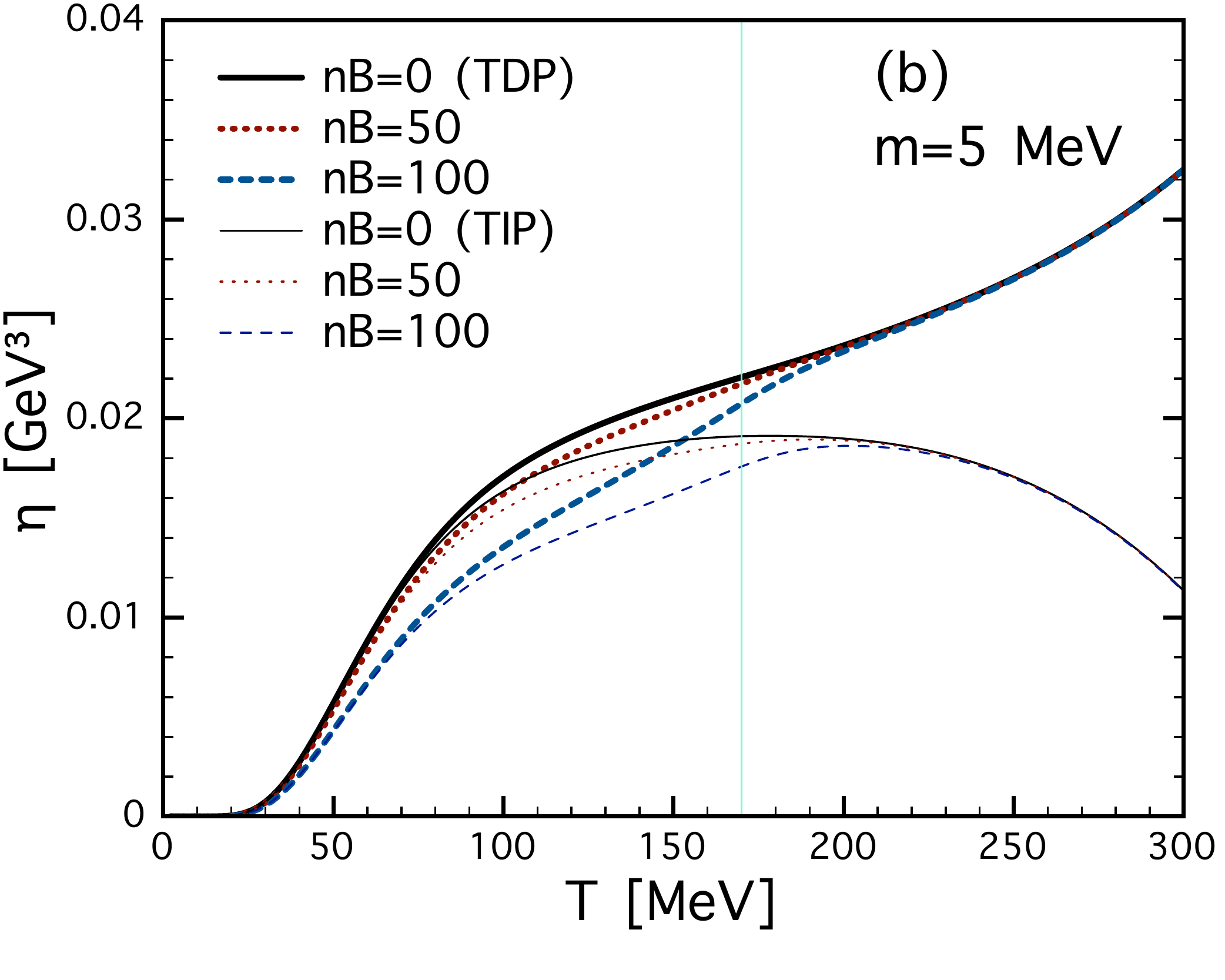}
\end{tabular}
\caption{(a) Shear viscosities $\eta$ in the chiral limit as functions of $T$ for different strengths for the {\it static} external magnetic field $eB_0=n_Bm^2_\pi$ for $n_B=0$ (solid), $50$ (dotted), and $100$ (dashed) with the $T$-dependent parameters (TDP, thick) and $T$-independent parameters (TIP, thin). (b) The same curves with $m=5$ MeV. The vertical lines indicate the chiral-phase-transition temperatures $T_0=(166,170)$ MeV for the (left,right) panels.}
\label{FIG23}
\end{figure}

In this Section, we present and discuss the numerical results of the shear viscosity. In Figure~\ref{FIG23}, the shear viscosity is presented as a functions of $T$ under the external magnetic field $B_0=n_B m^2_\pi$ in the chiral limit (a) and in the case of finite current-quark mass $m=5$ MeV (b). The thick and thin lines indicate those with $T$-dependent parameters (TDP) and $T$-independent parameters (TIP), respectively. Note that TDP stands for that the scale parameter $\bar{\rho}$ and $\bar{R}$ are functions of $T$ as shown in the left panel of Figure~\ref{FIG01}, whereas TIP means these two parameters are $T$-independent, i.e. $\bar{\rho}=\bar{\rho}(0)$ and the same for $\bar{R}$. The vertical lines shown in the both panels denote the transition temperatures $T_{0}$. 

We observe that the viscosity starts from zero, then keeps increasing as $T$ increases for TDP, whereas it decreases beyond $T_0$ for TIP. This observation suggests that the nontrivial $T$ dependence of the scale parameter of a nonperturbative model makes significant effects on the behavior of the shear viscosity.  As $T$ becomes higher, the scale parameter $\bar{\rho}$ decreases implying the reduction of the SB$\chi$S effect in our theoretical framework: The (anti)instanton ensemble is diluted as $T$ increases. In other words, the system is no longer strongly coupled one. It results in the larger shear viscosity. It is worth noting that the similar increasing behavior was also observed in the NJL model calculation~\cite{Fukutome:2007ta}. Although they considered small quark chemical potential $\mu=10$ MeV and they treat the finite width for the quark spectral function as a free parameter. 

Furthermore, the shear viscosity becomes smaller under the strong magnetic field for the both cases of $m=(0,5)$ MeV. Again, this tendency can be understood by the enhancement of SB$\chi$S, in terms of the magnetic catalysis~\cite{Miransky:2002rp}. In our theoretical framework, the magnetic field contributions are proportional to $\tilde{M}_{\bm{k}}\propto M_0(T)$ as shown in Eqs.~(\ref{eq:MMEE}) and (\ref{eq:FSV2}). They disappear beyond $T_0$ for the chiral limit as shown in the panel (a) of Figure~\ref{FIG23}, due to the second-order chiral phase transition as understood by seeing the panel (b) in Figure~\ref{FIG01}. the magnetic-field effects remain finite even beyond $T_0$ in the case of the finite current-quark mass as in the panel (b) of Figure~\ref{FIG23}. It is because of the crossover pattern of the chiral restoration.  At very high temperatures such as $T\gtrsim220$ MeV, the magnetic-field effects almost vanish even in the finite current-quark mass. Near the transition temperature $T_0\approx170$ MeV, the shear viscosity becomes approximately $\eta\approx0.02\,\mathrm{GeV}^3$ for all the cases.

\begin{figure}[t]
\begin{tabular}{cc}
\includegraphics[width=8.5cm]{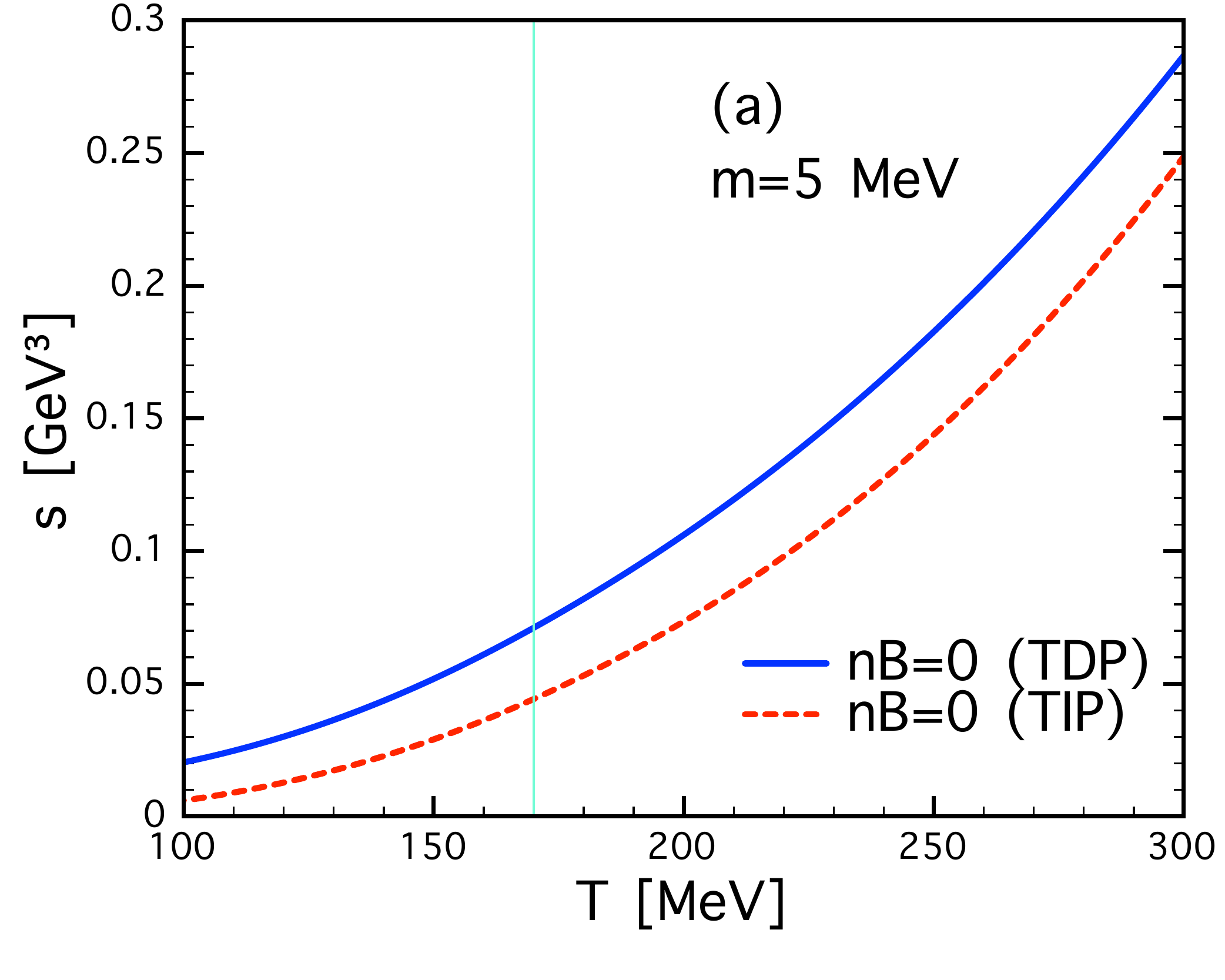}
\includegraphics[width=8.5cm]{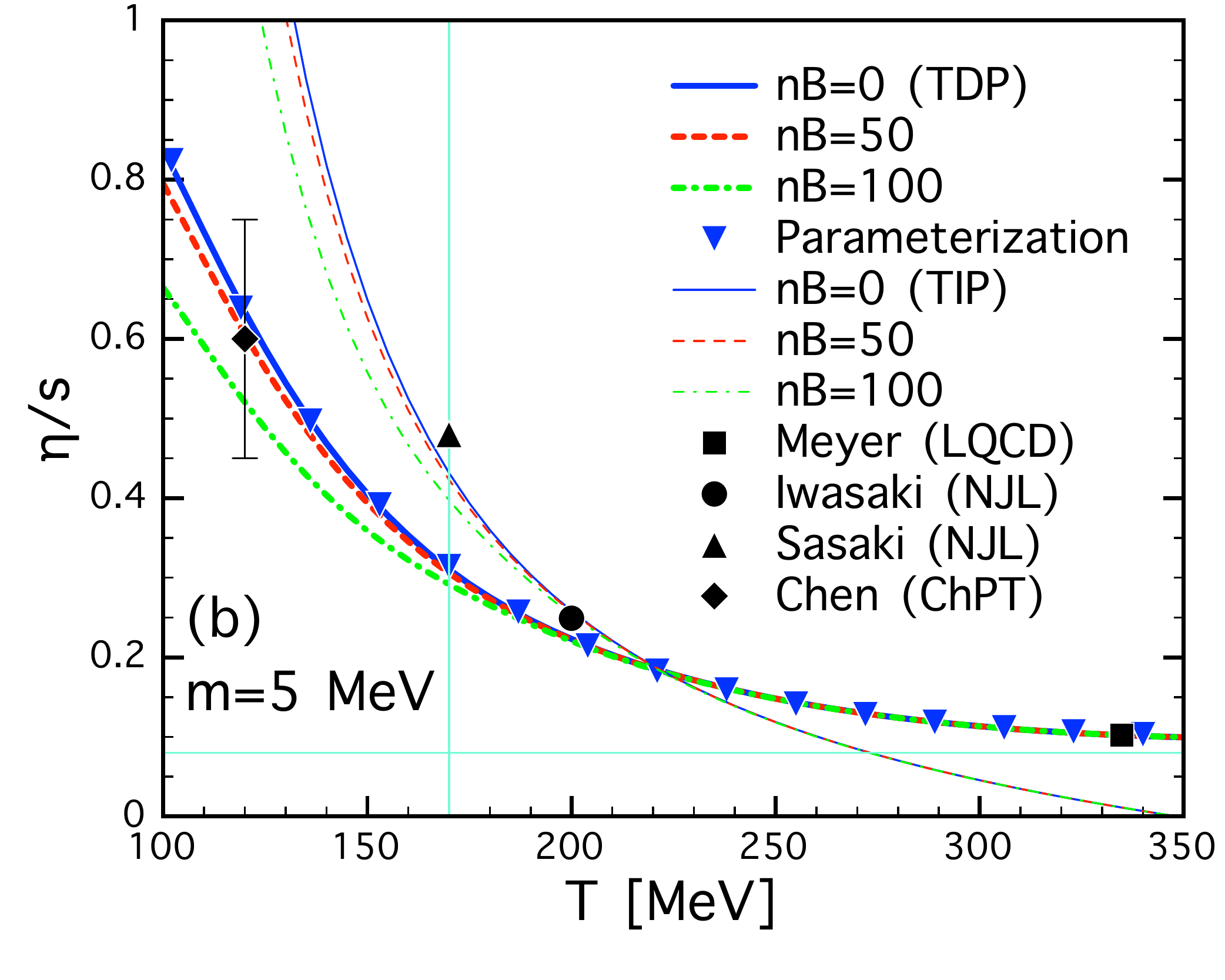}
\end{tabular}
\caption{(Color online) (a) Entropy density $s$ as a function of $T$ for with the $T$-dependent parameters (TDP, thick) and $T$-independent parameters (TIP, thin). (b) The ratio of the shear viscosity and entropy density $\eta/s$ in the same manner with the left panel. with different strengths of the external magnetic fields, $n_B=(0,50,100)$, given in the (solid, dash, dot-dash) lines. We also showed the theoretical results from Meyer (LQCD)~\cite{Meyer:2007ic} (square), Iwasaki (NJL)~\cite{Iwasaki:2007iv} (circle),  Sasaki (NJL)~\cite{Sasaki:2008um} (triangle), and Chen ($\chi$PT)~\cite{Chen:2006iga} (diamond). The parameterization of the TDP curve for $n_B=0$ in Eq.~(\ref{eq:PARAETA}) is also given with the solid nabla. Detailed explanations for these theoretical values are given in the text. The vertical lines indicate the chiral-phase-transition temperatures $T_0=170$ MeV for the (left,right) panels, while the horizontal one in the right panel stands for the lower bound of the QGP shear viscosity, i.e. the KSS bound $\eta/s=1/(4\pi)\approx0.08$.}
\label{FIG45}
\end{figure}

In literatures, the ratio of the shear viscosity and the entropy density $\eta/s$ has been considered as an important physical quantity. Hence we also present our result for $\eta/s$ here. First, in the left panel of Figure~\ref{FIG45}, we depict the entropy density using Eq.~(\ref{eq:ENT1}) for TDP (solid) and TIP (dash). Since we are interested only in the cases with the finite current-quark mass, we choose $m=5$ MeV, as mentioned above. We find that $\eta/s$ are smoothly increasing curves with respect to $T$ for the both cases. whereas the result for TDP is larger than the other. This can be easily understood by that the first term in the right-hand-side of Eq.~(\ref{eq:ENT1}) becomes zero for $N/V=1/\bar{R}^4=\mathrm{const.}$. Note that here we set the external magnetic field to be zero here, since we have verified that the magnetic-field contributions to the entropy density are negligible.

In the right panel of Figure~\ref{FIG45}, we show the numerical results for the ratio $\eta/s$ as functions of $T$ for TDP (thick) and TIP (thin), with different strengths of the magnetic field. 
Even the $T$ dependence of the model parameters have been taken into account, 
the present model scale is about $\Lambda\approx600$ MeV since it corresponds to the nonperturbative QCD region. Therefore, we confine our discussion to the temperature not much far beyond the chiral transition, i.e. $T_\mathrm{max}=350$ MeV. The magnetic field dependence of $\eta/s$ comes only from the numerator $\eta$. Again, here we choose the current-quark mass as $m=5$ MeV which manifests the crossover chiral restoration pattern. The horizontal and vertical lines stand for the chiral transition temperature $T_0=170$ MeV and the KSS-bound value $\eta/s=1/(4\pi)\approx0.08$, respectively. The curves of $\eta/s$ for TDP decrease smoothly and approaches to the KSS bound. Those for TIP behave similarly but decrease more stiffly with respect to $T$. Note that the TIP curves undershoot the KSS bound at $T\approx270$ MeV. These observations of the different curve behaviors implies that it is necessary to take the temperature dependence of the model parameters into consideration. The effects of the magnetic field are sizable below the chiral transition, then become negligible beyond $T_0$.  Near the transition point, we observe only a few percent changes in the ratio $\eta/s$ due to the magnetic field.

In the right panel of Figure~\ref{FIG45}, the other theoretical estimations for the ratio $\eta/s$ is also presented for comparison. In Ref.~\cite{Meyer:2007ic}, the Monte-Carlo calculations of the two-point correlations in the pure SU(3) gauge were employed to compute the ratio with the nonperturbatively normalized operators. It gives $\eta/s=(0.134,0.102)$ at $T=(1.65,1.24)\,T_c$, where $T_c$ is the critical temperature. This result is represented by solid square. The TDP curves are well compatible with their result at $T\approx335$ MeV, while the TIP curves undershoot the value.

The effective models such as the NJL model have been also used for estimating the ratio as well. In Ref.~\cite{Iwasaki:2007iv}, it has been reported  that $\eta/s\approx(1/4\pi\sim0.9)$ at $(T,\mu)=(200,10)$ MeV, depending on the finite width for the quark spectral function. Averaging their values over the finite width, we have $\eta/s\approx0.25$, and it is represented by the solid circle in the left panel of Figure~\ref{FIG45}. It locates between the TDP and TIP curves.
In their previous work with the same theoretical framework~\cite{Fukutome:2007ta}, the shear viscosity increases with the larger quark chemical potential with a small increasing rate. Hence the depicted point in the right panel is supposed to be lowered at $\mu$=0. Nevertheless, the change from $\mu=(10\to0)$ MeV will not substantial in the present discussions.

Employing the NJL model, Ref.~\cite{Sasaki:2008um} explored the transport coefficients near the chiral phase transition. From their results of the ratio $\eta/s\approx0.5$ at $T\approx170$ MeV. This value is depicted in the right panel of Figure~\ref{FIG45} with the solid triangle. It is comparable with the TIP curves and but larger compared with the TDP curves by about $50\%$.
Note that the $\eta/s$ curves in Ref.~\cite{Sasaki:2008um} show similar $T$-dependency to ours. However, beyond $T_0$, their curves turn into a slightly increasing ones and no longer similar to ours. In Ref.~\cite{Chen:2006iga} they computed $\eta/s$ by using $\chi$PT below the chiral transition temperature. They estimated $\eta/s$  as a decreasing function of $T$ with a typical value $\eta/s=0.6$ at $T=120$ MeV with $50\%$ uncertainty. We depict this value with the solid diamond with the error bar in the right panel of Figure~\ref{FIG45}, well-matching with the TDP curves. There are other theoretical estimations for $\eta/s$ for the high-$T$ ($T\gtrsim 450$ MeV) regions from LQCD and pQCD~\cite{Nakamura:2004sy,Chen:2009sm,Arnold:2003zc,Xu:2007ns}, and those results can not be reproduced within the present model: The theoretical results are usually larger than ours by $(5\sim10)$ times. However, this situation is rather natural, considering that the present model is well compatible in the low-energy regions as a nonperturbative effective model. Hence, we do not discuss those result here.

Finally, from the numerical results obtained above, we want to provide a simple parameterization of the ratio $\eta/s$ as a function of $T$. Since many theoretical approaches for the QGP dynamics have used a $T$-independent $\eta/s$ value~\cite{Song:2010mg}, this parameterization would help to construct more realistic models of QGP. Taking into account that the magnetic-field effects are negligible for $T>T_0$ as shown in the right panel of Figure~\ref{FIG45}, we just parameterize the numerical result for $n_B=0$. Employing a simple analytic form, one is led to
\begin{equation}
\label{eq:PARAETA}
\frac{\eta}{s}=0.27-\frac{0.87}{t}+\frac{1.19}{t^2}-\frac{0.28}{t^3},\,\,\,\,T=(100\sim350\,\mathrm{MeV}),
\end{equation}
where we use a notation $t=T/T_0$ with $T_0=170$ MeV. In the right panel of Figure~\ref{FIG45}, we show the $\eta/s$ curve using the parameterization in Eq.~(\ref{eq:PARA}) with the solid nabla.

\section{Summary, conclusion, and future perspectives}
In summary, we have investigated the QGP shear viscosity of the SU(2) light-flavor quark matter at finite temperature under the strong external magnetic field. We employed the liquid instanton model and Green-Kubo formula to derive our result. The external magnetic field has been induced by the Schwinger method. Since the shear viscosity becomes zero at the mean-field level, we suggested a finite-width quark spectral function motivated by the instanton model. There is one free parameter in this quark spectral function which has been determined by the chiral condensate value at zero temperature. The model parameters such as the average instanton size and inter-instanton distance are all temperature-dependent. This character is different from usual local-interaction models. We list our important observations as follows:
\begin{itemize}
\item In our model, many parameters such as quark mass and instanton size are modified by temperature. Our way of modifying those parameters are supported by the fact that our effective thermodynamic potential at the leading large-$N_c$ contributions manifests the correct chiral restoration patterns, i.e. the second order and crossover phase transitions for $m=0$ and $m\ne0$, respectively.
 \item Our study here has shown that the external magnetic field reduces $\eta$ due to the magnetic catalysis, i.e. the quarks are coupled more strongly in the presence of the magnetic field. The effect from the external magnetic field turns out to be sizable below the chiral transition temperature $T_0=(166,170)$ MeV for $m=(0,5)$ MeV. However it becomes inappreciable when temperature goes  beyond $T_0$
     since the nonperturbative effects, such as the magnetic catalysis, becomes diminished. We also find that a typical value for the shear viscosity near $T_0$ is $\eta=0.02\,\mathrm{GeV}^3$.
 \item We observe that the $T$-dependent parameters (TDP), $\bar{\rho}(T)$ and $\bar{R}(T)$ play an important role beyond $T_0$ which make $\eta$ to increase. In contrast, $\eta$ starts to decrease after $T_0$, if we choose the $T$-independent parameters (TIP).  The ratio of the shear viscosity and the entropy density, $\eta/s$ is also computed for the finite current-quark mass. It has been shown to be a monotonically decreasing function of $T=(100\sim350)$ MeV. Furthermore, we also find that $\eta/s$ undershoots the KSS bound, $\eta/s=1/(4\pi)$, for TIP. On the other hand, $\eta/s$ approaches to the KSS bound for TDP. At $T_0=170$ MeV, we find a typical value for the ratio as $\eta/s=0.29$ within the present model.
 \item Our numerical results of $\eta/s$ for TDP are well comparable with other theoretical estimations such as the NJL model,  LQCD, and $\chi$PT for $T=(100\sim350)$ MeV, although we can not reproduce the very high-$T$ results from LQCD and pQCD. However, this is rather natural, since the present model is well applicable for the low-energy regions. As for the future usage in QGP-dynamics studies, we also parameterize the numerical result of $\eta/s$ in a simple polynomial form as a function of $t=T/T_0$ without the magnetic field.
\end{itemize}

Encouraged by the fact our results obtained in the present work agree well with the empirical data,
we would like to extend our study to other QGP transport coefficients, such as the bulk viscosity and heat conductivity. Moreover, it would be interesting to take into account the external electric field, which turns out to be considerably strong in heavy-ion collisions. Thus, the external electric field may bring considerable changes in the transport coefficients, in comparison to the pure magnetic one.  Related works are under progress and will appear elsewhere.
\section*{Acknowledgment}
The authors thank  S.~Cho (Yonsei Univ.) and  L.~Wu (NCTU) for fruitful discussions. S.i.N. is grateful to the warm hospitality during his stay at National Center for Theoretical Science (NCTS, north) of Taiwan, where the present work was partially performed. He also acknowledges the financial supports from NCTS for his stay. The work of C.W.K. was supported by the grant NSC 99-2112-M-033-004-MY3 from National Science Council (NSC) of Taiwan. He has also acknowledged the support of NCTS (North) of Taiwan. The numerical calculations were carried out in part using ABACUS2 computing sever at KIAS.

\section*{Appendix}
The quark spectral function is normalized as follows:
\begin{eqnarray}
\label{eq:COR}
&&\frac{1}{2\pi}\int^\infty_{-\infty}\rho_\mathrm{FW}(w,\bm{k})dw=
\int^\infty_{-\infty}
\frac{\mathrm{sgn}(w)\,(w\gamma_0-\alpha)}{2\sqrt{2\pi}E\Lambda}\left[\exp\left[-\frac{(w-E)^2}{2\Lambda^2} \right]+\exp\left[-\frac{(w+E)^2}{2\Lambda^2} \right] \right]dw
\cr
&=&\int^\infty_{-\infty}
\frac{\mathrm{sgn}(w)\,(w\gamma_0-\alpha)}{2\sqrt{2\pi}E\Lambda}\left[\exp\left[-\frac{(w-E)^2}{2\Lambda^2} \right]+\exp\left[-\frac{(w+E)^2}{2\Lambda^2} \right] \right]dw.
\end{eqnarray}
Replacing the integral variable as $w\pm E\equiv w_\pm$, Eq.~(\ref{eq:COR}) is led to
\begin{eqnarray}
\label{eq:COR2}
&&\int^\infty_{-\infty}\left\{
\frac{\mathrm{sgn}(w_+-E)\,[(w_+-E)\gamma_0-\alpha]}{2\sqrt{2\pi}E\Lambda}
\exp\left[-\frac{w_+^2}{2\Lambda^2} \right]+
\frac{\mathrm{sgn}(w_-+E)\,[(w_-+E)\gamma_0-\alpha]}{2\sqrt{2\pi}E\Lambda}\exp\left[-\frac{w_-^2}{2\Lambda^2} \right]\right\}dw
\cr
&&=\frac{\mathrm{sgn}(-E)(-E\gamma_0-\alpha)}{2E}
+\frac{\mathrm{sgn}(E)(E\gamma_0-\alpha)}{2E}
=\frac{-(-E\gamma_0-\alpha)}{2E}
+\frac{+(E\gamma_0-\alpha)}{2E}=\gamma_0,
\end{eqnarray}
which satisfies the spectral function normalization condition.



\begin{thebibliography}{99}
\bibitem{Song:2010mg}
  H.~Song {\it et al},
  Phys.\ Rev.\ Lett.\  {\bf 106}, 192301 (2011)
  [Erratum-ibid.\  {\bf 109}, 139904 (2012)].
\bibitem{Policastro:2001yc}
  G.~Policastro, D.~T.~Son and A.~O.~Starinets,
  Phys.\ Rev.\ Lett.\  {\bf 87}, 081601 (2001).
\bibitem{Buchel:2003tz}
  A.~Buchel and J.~T.~Liu,
  Phys.\ Rev.\ Lett.\  {\bf 93}, 090602 (2004).
\bibitem{Kovtun:2004de}
  P.~Kovtun, D.~T.~Son and A.~O.~Starinets,
  Phys.\ Rev.\ Lett.\  {\bf 94}, 111601 (2005).
\bibitem{Arsene:2004fa}
  I.~Arsene {\it et al.}  [BRAHMS Collaboration],
  Nucl.\ Phys.\ A {\bf 757}, 1 (2005).
\bibitem{Abelev:2008ab}
  B.~I.~Abelev {\it et al.}  [STAR Collaboration],
  Phys.\ Rev.\ C {\bf 79}, 034909 (2009).
\bibitem{Schenke:2012wb}
  B.~Schenke, P.~Tribedy and R.~Venugopalan,
  Phys.\ Rev.\ Lett.\  {\bf 108}, 252301 (2012).
\bibitem{Song:2012ua}
  H.~Song,
  arXiv:1210.5778 [nucl-th].
\bibitem{Cremonini:2011iq}
  S.~Cremonini,
  Mod.\ Phys.\ Lett.\ B {\bf 25}, 1867 (2011).
\bibitem{Carrington:1999bw}
  M.~E.~Carrington, D.~f.~Hou and R.~Kobes,
  Phys.\ Rev.\  D {\bf 62}, 025010 (2000).
\bibitem{Fukutome:2007ta}
  T.~Fukutome and M.~Iwasaki,
  Prog.\ Theor.\ Phys.\  {\bf 119}, 991 (2008).
\bibitem{Chen:2006iga}
  J.~W.~Chen and E.~Nakano,
  Phys.\ Lett.\  B {\bf 647}, 371 (2007).
\bibitem{Chen:2009sm}
  J.~W.~Chen, H.~Dong, K.~Ohnishi and Q.~Wang,
  Phys.\ Lett.\  B {\bf 685}, 277 (2010).
\bibitem{Arnold:2003zc}
  P.~B.~Arnold, G.~D.~Moore and L.~G.~Yaffe,
  JHEP {\bf 0305}, 051 (2003).
\bibitem{Arnold:2000dr}
  P.~B.~Arnold, G.~D.~Moore and L.~G.~Yaffe,
  JHEP {\bf 0011}, 001 (2000).
\bibitem{Huang:2009ue}
  X.~G.~Huang, M.~Huang, D.~H.~Rischke and A.~Sedrakian,
  Phys.\ Rev.\  D {\bf 81}, 045015 (2010).
\bibitem{Iwasaki:2007iv}
  M.~Iwasaki, H.~Ohnishi and T.~Fukutome,
  hep-ph/0703271.
\bibitem{Huang:2011ez}
  X.~G.~Huang and T.~Koide,
  Nucl.\ Phys.\ A {\bf 889}, 73 (2012).
\bibitem{Sasaki:2008um}
  C.~Sasaki and K.~Redlich,
  Nucl.\ Phys.\ A {\bf 832}, 62 (2010).
\bibitem{Green-Kubo:1957mj}
  R.~Green-Kubo,
  J.\ Phys.\ Soc.\ Jap.\  {\bf 12}, 570 (1957).
\bibitem{Meyer:2007ic}
  H.~B.~Meyer,
  Phys.\ Rev.\ D {\bf 76}, 101701 (2007).
\bibitem{Mamo:2012sy}
  K.~A.~Mamo,
  JHEP {\bf 1210}, 070 (2012).
\bibitem{Bzdak:2011yy}
  A.~Bzdak and V.~Skokov,
  Phys.\ Lett.\ B {\bf 710}, 171 (2012).
\bibitem{Tuchin:2013ie}
  K.~Tuchin,
  arXiv:1301.0099 [hep-ph].
\bibitem{Fukushima:2008xe}
  K.~Fukushima, D.~E.~Kharzeev and H.~J.~Warringa,
  Phys.\ Rev.\  D {\bf 78}, 074033 (2008).
\bibitem{Boomsma:2009yk}
  J.~K.~Boomsma and D.~Boer,
  Phys.\ Rev.\ D {\bf 81}, 074005 (2010).
\bibitem{Diakonov:2002fq}
  D.~Diakonov,
  Prog.\ Part.\ Nucl.\ Phys.\  {\bf 51}, 173 (2003).
\bibitem{Schafer:1995pz}
  T.~Schafer and E.~V.~Shuryak,
  Phys.\ Rev.\  D {\bf 53}, 6522 (1996).
\bibitem{Harrington:1976dj}
  B.~J.~Harrington and H.~K.~Shepard,
  Nucl.\ Phys.\  B {\bf 124}, 409 (1977).
\bibitem{Diakonov:1988my}
  D.~Diakonov and A.~D.~Mirlin,
  Phys.\ Lett.\  B {\bf 203}, 299 (1988).
\bibitem{Nam:2009nn}
  S.~i.~Nam,
  J.\ Phys.\ G {\bf 37}, 075002 (2010).
\bibitem{Fukushima:2003fw}
  K.~Fukushima,
  Phys.\ Lett.\  B {\bf 591}, 277 (2004).
\bibitem{Ratti:2005jh}
  C.~Ratti, M.~A.~Thaler and W.~Weise,
  Phys.\ Rev.\  D {\bf 73}, 014019 (2006).
\bibitem{Schwinger:1951nm}
  J.~S.~Schwinger,
  Phys.\ Rev.\  {\bf 82}, 664 (1951).
\bibitem{Nieves:2006xp}
  J.~F.~Nieves and P.~B.~Pal,
  Phys.\ Rev.\  D {\bf 73}, 105003 (2006).
\bibitem{Nam:2008ff}
  S.~i.~Nam, H.~Y.~Ryu, M.~M.~Musakhanov and H.~-Ch.~Kim,
  J.\ Korean Phys.\ Soc.\  {\bf 55}, 429 (2009).
\bibitem{Espinosa:2005gq}
  O.~Espinosa,
  Phys.\ Rev.\ D {\bf 71}, 065009 (2005).
\bibitem{Nam:2012sg}
  S.~i.~Nam,
  Phys.\ Rev.\ D {\bf 86}, 033014 (2012)
\bibitem{Nam:2011vn}
  S.~i.~Nam and C.~W.~Kao,
  Phys.\ Rev.\ D {\bf 83}, 096009 (2011).
\bibitem{Schafer:1996wv}
  T.~Schafer and E.~V.~Shuryak,
  Rev.\ Mod.\ Phys.\  {\bf 70}, 323 (1998).
\bibitem{Goeke:2007bj}
  K.~Goeke, M.~M.~Musakhanov and M.~Siddikov,
  Phys.\ Rev.\ D {\bf 76}, 076007 (2007)
\bibitem{sinam}
S.~i.~Nam, in preparation.
\bibitem{Miransky:2002rp}
  V.~A.~Miransky and I.~A.~Shovkovy,
  Phys.\ Rev.\  D {\bf 66}, 045006 (2002).
\bibitem{Nakamura:2004sy}
  A.~Nakamura and S.~Sakai,
  Phys.\ Rev.\ Lett.\  {\bf 94}, 072305 (2005).
\bibitem{Xu:2007ns}
  Z.~Xu and C.~Greiner,
  Phys.\ Rev.\ Lett.\  {\bf 100}, 172301 (2008).

\end{thebibliography}
\end{document}